\def\units#1{\hbox{$\,{\rm {#1}}$}}

\documentstyle[12pt,epsfig]{article}
\input{epsfig.sty}
\begin{document}
~
\vskip -2.0 cm
{\hskip 10.0 cm \bf DFUB 98/30}
\begin{center}
\vskip 1.0 cm
{\bf MAGNETIC MONOPOLES}
 \vskip .5 cm 
{G. GIACOMELLI and L. PATRIZII}
\vskip .7cm
{\it Dipartimento di Fisica dell'Universit\`a di Bologna and INFN, 
Sezione di Bologna
Viale Berti-Pichat 6/2, I-40127 Bologna, Italy }

{\it E-mail: giacomelli@bo.infn.it, patrizii@bo.infn.it}

\vskip .3 cm

{\small Lecture at the Fifth School on Non~-~Accelerator 
Particle Astrophysics\\
Trieste, 29 June~-~10 July 1998}
\vskip 1.5 cm
\end{center}  
\begin{abstract}

The main aim of this lecture is to give an update on the lectures on Magnetic 
Monopoles (MMs) of the 1995 Trieste School. The update concerns theoretical 
developments, searches for GUT monopoles in the penetrating cosmic radiation at 
ever lower fluxes, study of the energy losses of MMs in the Earth and 
in detectors, 
searches for low mass monopoles, and finally mention byproducts of MM 
searches.
\end{abstract}

\section {Introduction}

Several theories and many models predict the existence of magnetic 
mono\-poles 
(MMs) [1-5] with a magnetic charge which obeys the Dirac quantization rule 
[6], 
$eg=n\hbar c/2$, where $n$ is an integer, $n$=1,2,3...  For $n$=1, 
assuming that the basic electric 
charge is that of the electron, one has 
$g=g_{D}=\hbar c/2e = 3.29 \cdot 10^{-8}$ u.e.s.
Most 
theoretical papers concern superheavy magnetic monopoles predicted by Grand 
Unified Theories (GUT) of electroweak and strong interactions; there are 
also many 
papers concerning still heavier MMs, and others concerning MMs with intermediate 
 and  low masses [5]; if the basic electric charge is 
that of the 
quark $d$ the basic magnetic charge increases by a factor of three.

A MM and an atomic nucleus N may form a bound system. Some models 
predict the existence of states with both electric and magnetic charges 
(dyons) [7]. 
As far as the energy losses in matter, a dyon and a $(MM~+~N)_{bound}$
system behave in 
the same way. The system $(MM~+~N)_{bound}$ can be produced mainly via 
radiative capture [8],
$MM~+~N \to (MM~+~N)_{bound}~+~\gamma$,  
and the bound system may be subject to 
photodissociation [8], 
$(MM~+~p)_{bound}+~\gamma \to MM~+~p$.
Since the cross section for radiative capture is 
$1 <\sigma<10~mb $ for MMs with $10^{-4}<\beta<10^{-3}$,
it is possible that many MMs [9] in the cosmic radiation can be in the 
state  $(MM+p)_{bound}$.
Neglecting the catalysis of proton decay, the MMs or the $(MM+p)_{bound}$
systems with $\beta \geq 10^{-3}$
traversing rock, probably reach an equilibrium 
mixture of about $50\%$ MMs and $50\%$ $(MM+p)_{bound}$ [8]. 
The minimum velocity of the  $(MM+p)_{bound}$
system for which one may have a break-up reaction is $\beta \sim 10^{-3}$
( at this value the three particles in the final state (MM, proton and target 
nucleus) are essentially at rest in the center of mass frame of the MM).

Most experimental searches concern superheavy GUT magnetic monopoles and
most of the published flux limits apply to an isotropic flux of bare 
monopoles of unit 
magnetic charge (the monopole catalyzed nucleon decay is not considered). 
In order 
to search for MMs and dyons over a large $\beta$-range it is necessary to know 
how these 
particles loose energy in different types of detectors [9] and if they can 
cross the Earth [10].
There has been increasing interest in lower mass monopoles, such as those 
discussed in Refs. [5,12,13]. Proposals have been made to search for them at 
sea level 
and at high altitude [13] and also at the future LHC collider at CERN [14].

This lecture is an update of the lectures on MMs at the 1995 Trieste 4$^{th}$ 
School on Non 
Accelerator Particle Astrophysics [15]. Here there will be a more complete analysis of the
 energy 
losses of MMs and of dyons in different detectors and in the Earth; an 
update is 
made of the GUT MM searches. An extensive bibliography on MMs is given in Ref. 
[16]. As a byproduct we shall discuss the searches  and the present limits 
for
nuclearites. These limits are also valid for charged Q-balls, aggregates of 
supersymmetric particles [17].

\section{Energy losses of monopoles and dyons in detectors}
	One has to consider the total energy losses of MMs and of dyons in the 
detectors and also that fraction of the energy loss which leads to detection. 
In streamer 
tubes this fraction is the ionization energy loss in the gas, 
in scintillators it is the 
excitation energy loss which leads to the emission of light, in nuclear track 
detectors it is 
the restricted energy loss, t. i. the energy loss contained in a 10 nm 
diameter around 
the MM trajectory. In Ref. [9] a thorough analysis is made of these losses and 
of 
their dependence on the MM velocity v=$\beta$c.

\subsection{Total energy losses in detectors}

Let us consider first the scintillators and the streamer tubes. 
At {\em high velocity ($\beta >0.05$) }
the total energy losses of MMs in scintillators and streamer tubes 
are mainly due to ionization; they may be calculated using the Bethe-Bloch 
formula. 
At {\em  intermediate $\beta$} (few $10^{-3} < \beta < 10^{-2}$) 
the main contribution to energy losses is 
again due to ionization, and one uses the formula of Ahlen-Kinoshita [18] for 
nonconductors, corrected by Ritson [19]:

\begin{equation}
\left(\frac{dE}{dx}\right)_{\mbox{ionization}} \simeq 2.6 \cdot 10^{6} \left(
\frac{g}{g_{D}} \right)^{2} \beta^{1.7} \;\;\;\;\;\; \frac{\mbox{MeV}~\mbox{g}}
{\mbox{cm}^{2}}
\end{equation}

The results are shown as curves A in Fig.~1a for MMs in scintillators.		

At {\em low velocity} ($10^{-5} < \beta < 10^{-3}$) the dominant energy 
loss mechanism is that of elastic collisions of the MM 
with the atoms. 
This contribution may be computed by 
numeric procedures, neglecting the hydrogen atoms (which means 
that one obtains a lower limit for the energy losses). The results are  
shown as curves B in Fig.~1a for MMs in scintillators [9].

\subsection{Light yields in scintillators for MMs}

For MMs the light yield $dL/dx$ in a scintillator is related to 
the energy loss $dE/dx$ by [21]
\begin{equation}
\frac{dL}{dx} = A \cdot \left[ \frac{1-F}{1+A
B(1-F)(dE/dx)}+F \right] \cdot
\frac{dE}{dx}
\label{Yield}
\end{equation}
where $dE/dx$ is the total electronic energy loss. For relatively small 
$dE/dx$ the light yield is proportional to the energy loss with the
 proportionality constant $A$. $B$ is the 
the so-called quenching parameter: the light yield from the energy deposited 
near the 
track (the first term in the parenthesis of Eq. 2) saturates for large 
energy losses. For $\beta > 0.1$
some electrons have sufficient energy 
to escape from the region near the track core ($\delta$ rays); $F$ is the fraction 
of energy loss 
which results from excitations outside the core; 
these excitations are assumed not to be quenched.

\begin{figure}[ht]
\vspace{-1cm}
\begin{center}
        \mbox{ \epsfysize=8.5cm
            \epsffile{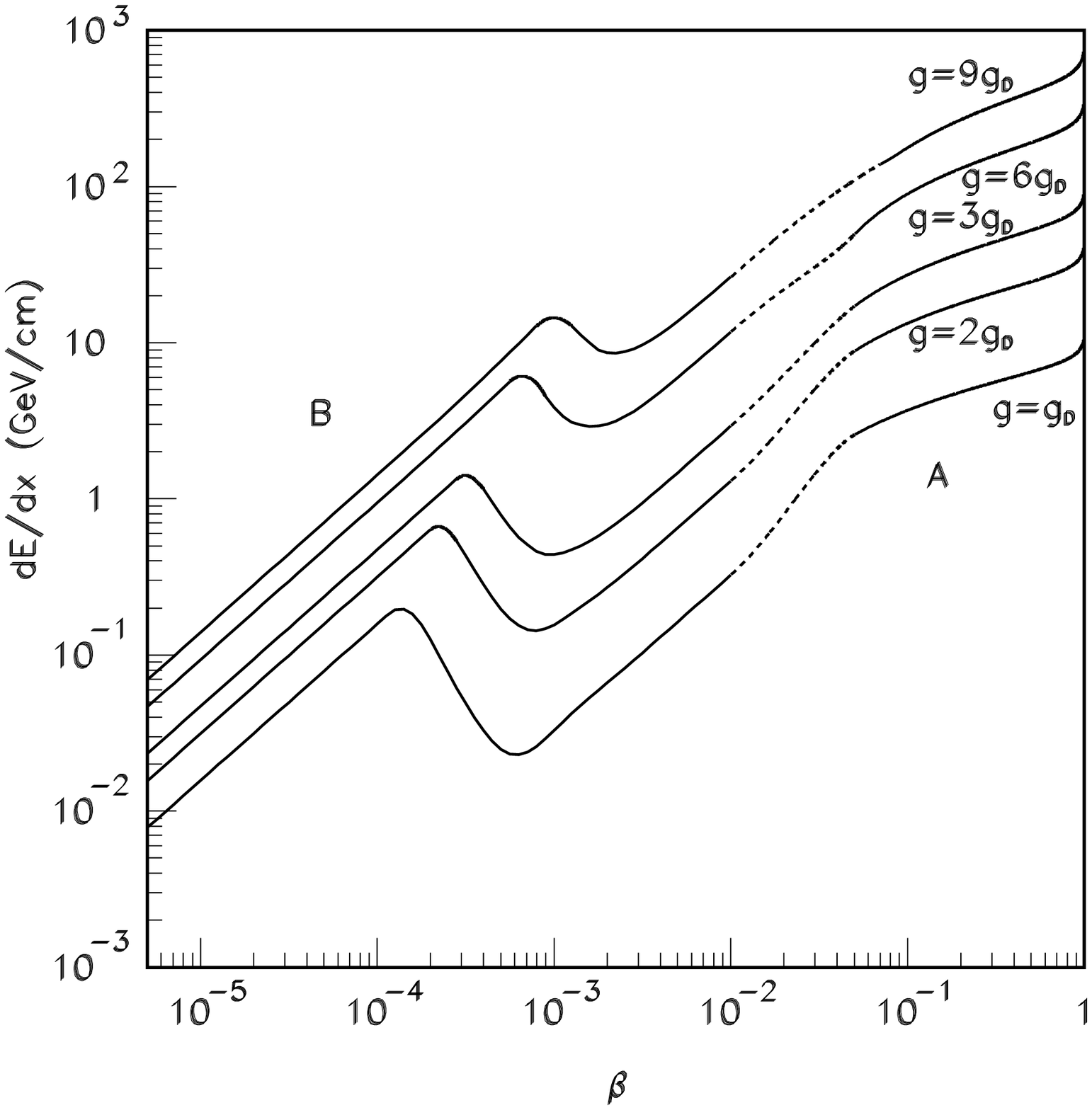}
            \hspace{1cm}
             \epsfysize=8.5cm
            \epsffile{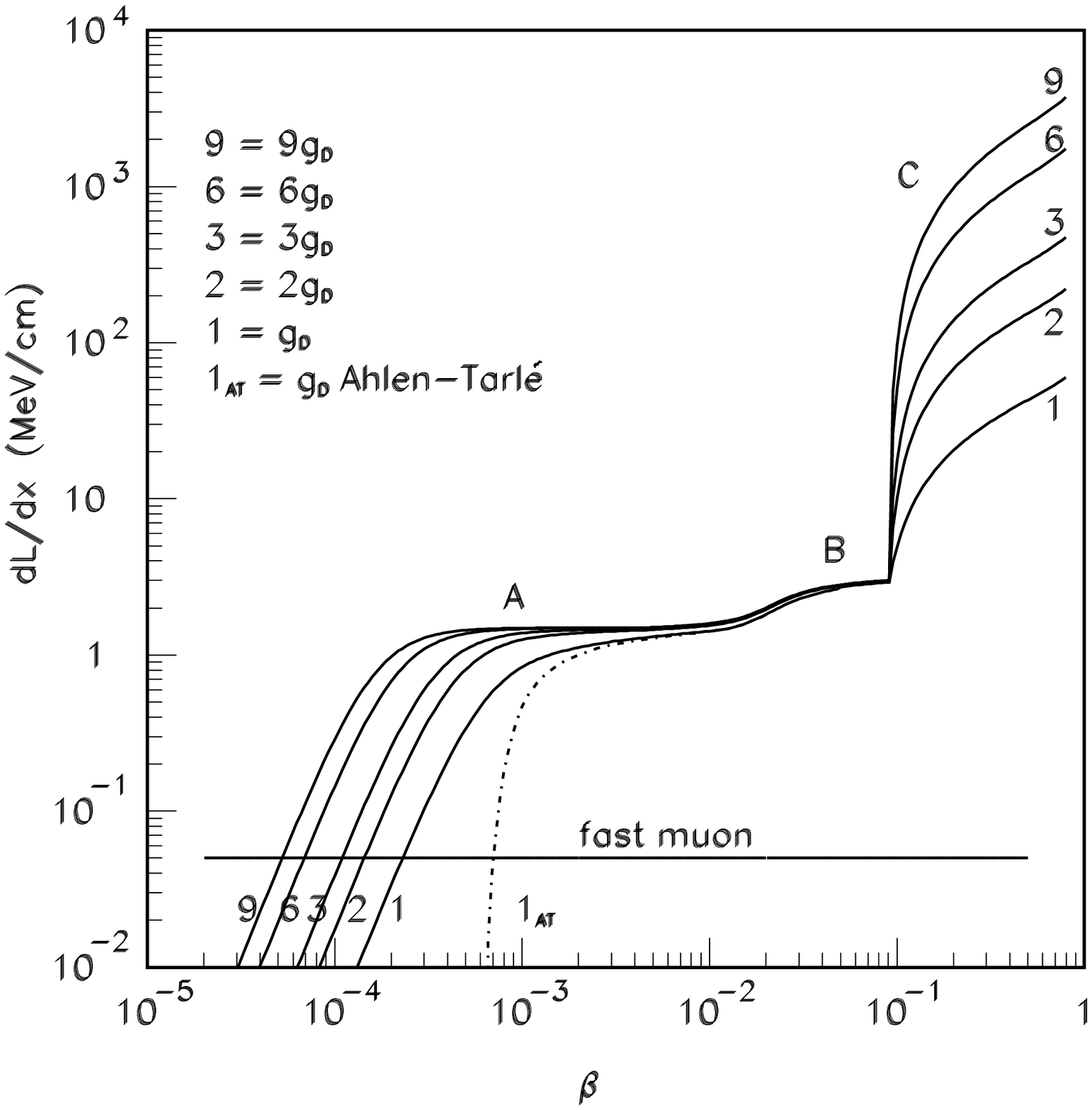}}
\vspace{-3.0cm}
\begin {center}
{\small \hspace{4cm} (a) \hspace{7cm} (b) }
\end{center}
\vspace{1cm}
\caption{\protect\small (a) Total energy losses of MMs in the MACRO liquid
scintillator, as a function of $\beta=v/c$ for various magnetic
charges
$g$. The dashed lines indicate interpolations.
See text for the explanation of curves in regions A and  B.
(b) Light yield of MMs in the plastic scintillator NE110
($\rho = 1.032 \;\; \mbox{g}/\mbox{cm}^{3}$) and in the MACRO liquid
scintillator ($\rho = 0.86 \;\; \mbox{g}/\mbox{cm}^{3}$),
versus $\beta$ for different magnetic charges $g$.
No significant difference is
found between the losses in the two scintillators.}
\label{TELsc}
\end{center}
\end{figure}

 The energy loss $dE/dx$ for a monopole is given at {\em high velocity} 
($\beta > 0.05$)) by the corrected Bethe-Bloch formula.

At {\em  lower velocities} ($\beta < 0.01$) 
the energy losses of a MM in scintillators are
 $\frac{1}{4}(g/g_{D})^{2}$ times the energy losses of a proton 
of the same $\beta$; these are 
computed from the energy losses in hydrogen and carbon, adding an exponential 
factor from the fit to the low-velocity proton data. For 
$0.01 < \beta < 0.05$ a smooth interpolation is used.
The results of the calculations are shown in Fig.~1b. Notice that at 
{\em very low velocities} ($ 0.0002<\beta<0.0005$ for $g=g_{D}$)
the light yield increases with $\beta$, then it saturates at a 
value of 1.2 MeV/cm (region A of Fig.~1b). The increase in the light yield 
observed in 
region B of Fig.~1b, is due to changes in the quenching parameters. For 
$\beta \ge 0.09$ there is
production of delta rays, and the light yield increases 
again with $\beta$ (region C of Fig.~1b).

For $0.003< \beta < 0.1$ 
{\it the light yield is thus independent of 
the magnetic charge value}; at low and high $\beta$ 
the light yield increases quadratically 
with $g$. The light yields in NE110 and in the MACRO liquid scintillator 
are equal 
within a few percent and only one curve is drawn.  The light yield at 
low beta is a lower limit, since the present calculations do not take into 
account possible contributions that  arise from the mixing and crossing 
(Drell effect) of 
molecular electronic energy levels at the passage of the magnetic charge; 
this could 
result in molecular excitations and emission of light.

The Ahlen-Tarl\'{e} curve shown in Fig.~1b, refers to the calculation of 
Ref.~[22] 
which predicts a kinematical cut off at $\beta = 7\cdot 10^{-4}$
due to the excitation energy of benzene 
molecules (5 eV); however experiments with low energy protons [23] have shown 
that no such cut off occurs.
The light yield is a small fraction of the total energy loss, see Figs.~1a 
and 1b.

\subsection{Energy losses of MMs in streamer tubes}

At {\em high velocities}, $\beta > 0.05$, the Bethe Bloch formula holds for 
the energy losses of MMs in the streamer tubes. The resulting energy losses 
are given in Fig.~2a, as curves A.\par
At {\em low velocities,} $10^{-4}< \beta <10^{-3}$,
and for $g=g_{D}$ in helium,  the Drell effect 
occurs: the energy levels of the helium atoms are changed by the presence of a 
magnetic charge, and at the passage of the magnetic charge an electron 
can make a transition to an excited level:
 
\begin{equation}
\left(\frac{dE}{dx}\right)_{Drell} = 11(\beta /10^{-4})[1-(9.3 \cdot
10^{-5}/\beta)^{2}]^{3/2}
\;\;\;\mbox{MeV cm}^{2}\mbox{/g}
\end{equation}

\noindent Such energy losses lead to the atomic excitation of helium atoms; these lead in 
turn to the ionization of the n-pentane molecules via the Penning effect.

At {\em intermediate velocities,} $2 \cdot 10^{-3} < \beta < 7
\cdot 10^{-2}$,
one considers the medium as a degenerate electron gas.
The results of the calculations are shown in Fig.~2a, as curves B. 

For dyons one has to sum up the direct excitation and ionization produced 
by the electric charge and 
by the moving magnetic charge. In addition the elastically recoiling hydrogen 
and carbon nuclei add contributions.
The results of the calculation for the energy losses of dyons are shown in 
Fig.~2b.

\begin{figure}[ht]
\vspace{-1cm}
\begin{center}
        \mbox{ \epsfysize=8.5cm
                \epsffile{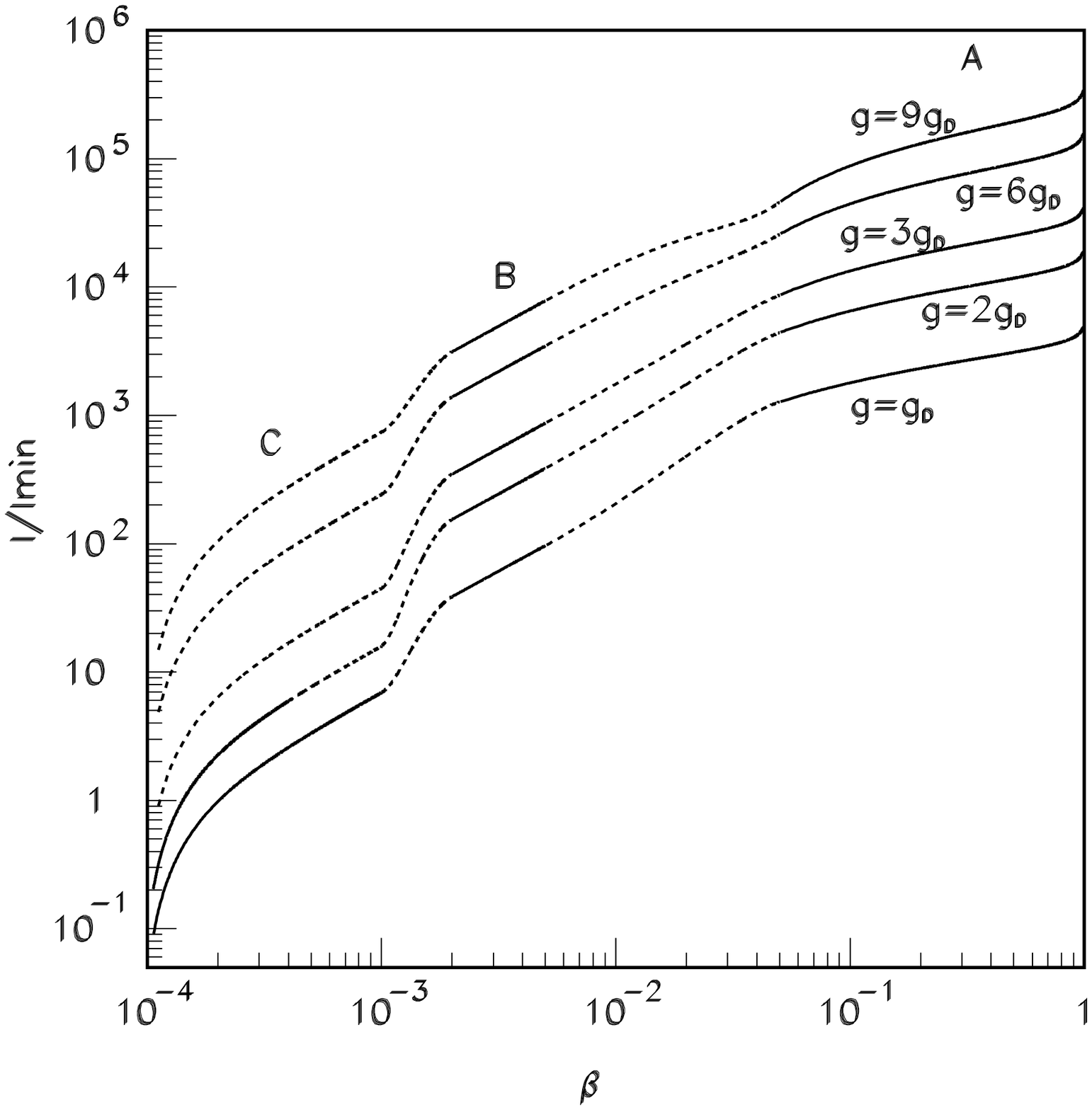}\hspace{1cm}
		\epsfysize=8.5cm
		\epsffile{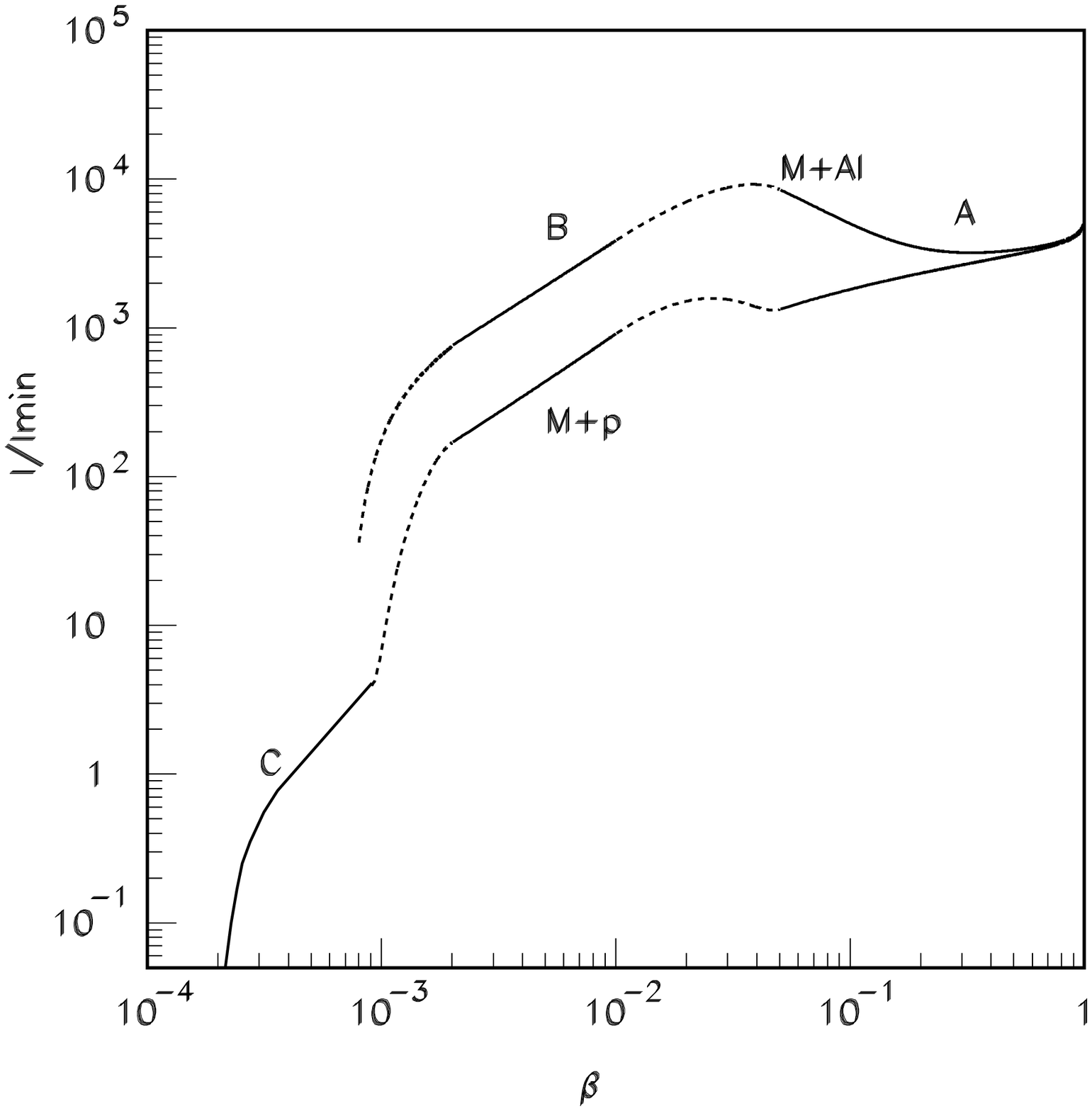}}
\vspace{-3.0cm}
\begin {center}
{\small \hspace{4.3cm} (a) \hspace{6.5cm} (b) }
\end{center}
\vspace{1cm}
\end{center}
{\small Figure 2: Energy losses in the 
limited streamer tubes filled with
73\% He, 27\%
n-pentane, $\rho = 0.856~~ \mbox{mg}/\mbox{cm}^{3}$,
versus $\beta$, relative to the
ionization produced by a minimum ionizing particle,
$I_{min} = 2.2~~\mbox{MeV g/cm}^{2}$: (a) by MMs, (b) by dyons with $g=g_{D}$ and $Q=+e$ (or by a M+p composite) and
$g=g_{D}$ and $Q=+13\; e$ (or by a M+Al composite).
The solid curves represent regions
where the calculations are more reliable; the dashed lines are
interpolated values [9].}
\label{MMion}
\end{figure}

\subsection{Energy losses of MMs and of dyons in the CR39 nuclear track 
detector}

The restricted energy losses of MMs and of dyons as a function of $\beta$ 
in the nuclear track detetctor CR39 have been computed using similar 
approximations and are presented in Figs. 3a and 3b.

\begin{figure}[ht]
\vspace{-2cm}
\begin{center}
        \mbox{ \epsfysize=8.5cm
            \epsffile{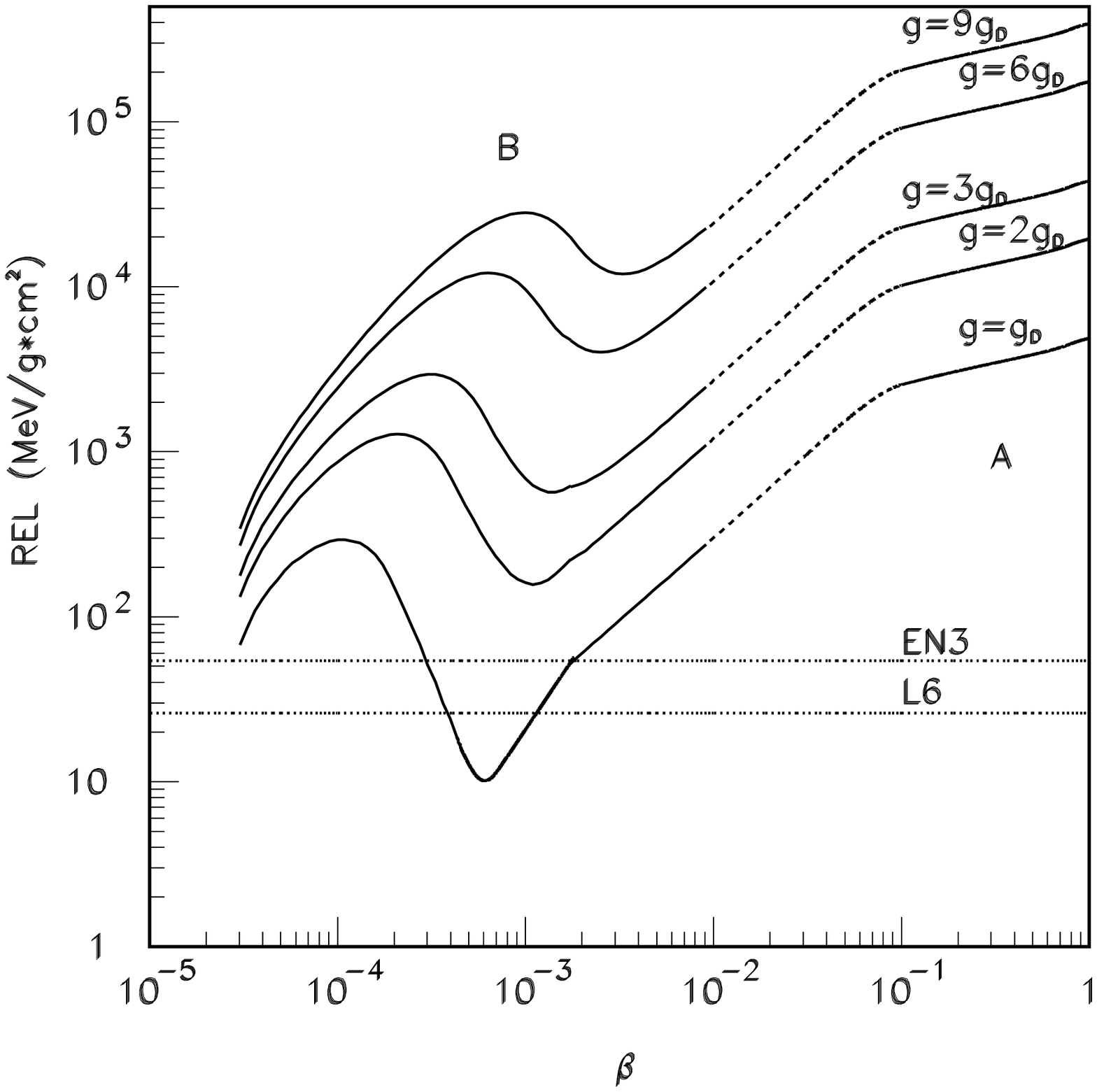} \hspace{1cm}
	\epsfysize=8.5cm
            \epsffile{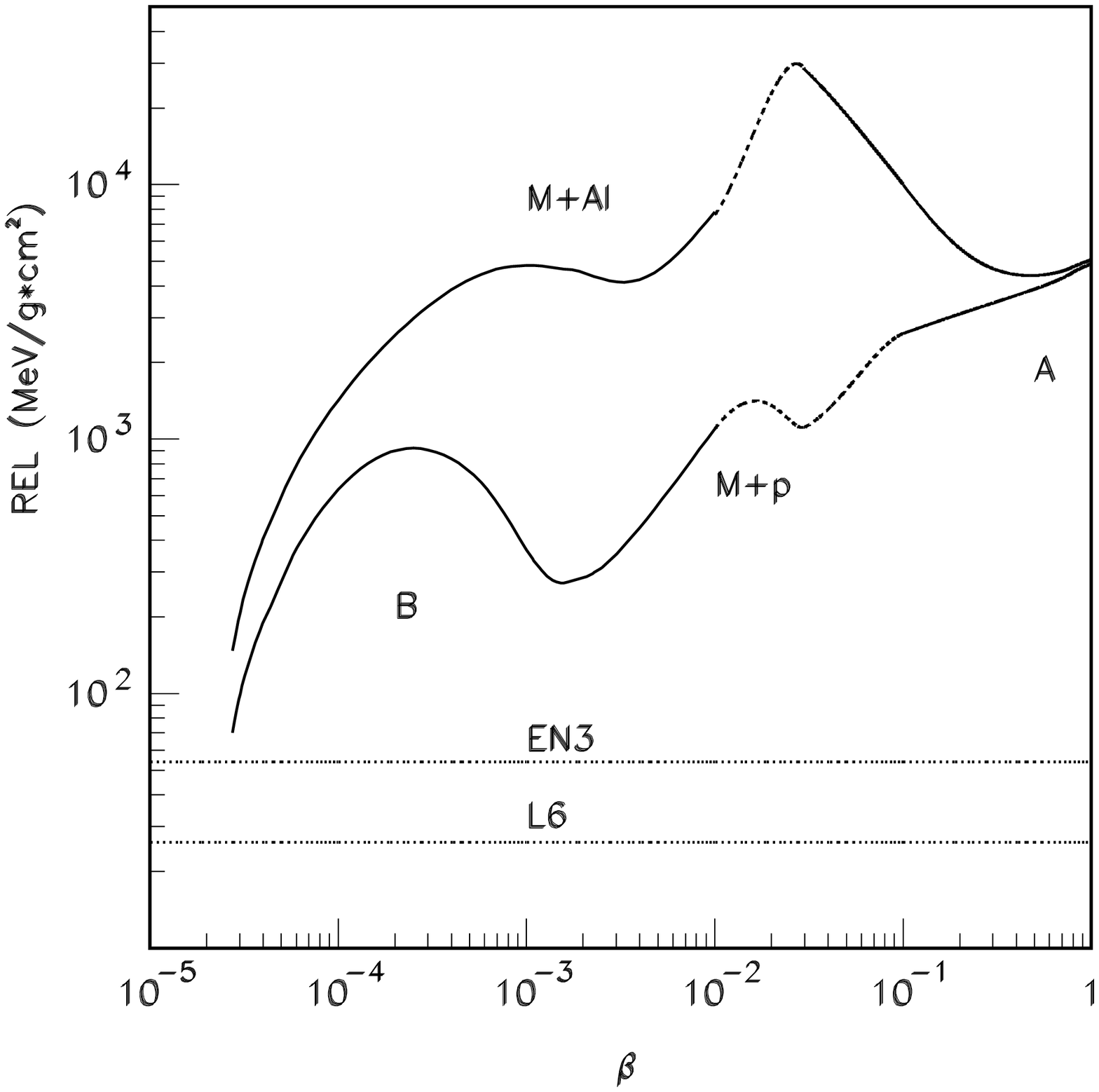}}

\vspace{-3.0cm}
\begin {center}
{\small \hspace{4.5cm} (a) \hspace{6.5cm} (b) }
\end{center}
\vspace{1cm}
\end{center}
{\small Figure 3: Restricted Energy Losses versus $\beta$ 
(a) for MMs, (b) for dyons with $g=g_{D}$ and $Q=+e$
(or for a M+p composite) or
$g=g_{D}$ and $Q=+13\; e$ (or for a M+Al composite) in the
nuclear track detector CR39 ($\rho = 1.31 \;\;\mbox{g}/\mbox{cm}^{3}$).
The detection thresholds of two types of CR39 used by the MACRO experiment
(EN3 and L6) are also shown [9]. See notes in Figures 1-2.}
\label{MMREL}
\end{figure}

\section{ Energy losses of MMs and of Dyons in the Earth.}

In order to compute the acceptance of a detector to MMs coming from above and 
from below one needs a model of the Earth mantle and nucleus, and an evaluation of 
the energy losses down to very small velocities.
The computations of Ref.~[10] show that only heavy monopoles can traverse the 
Earth: for istance for $g=g_D$ and  $\beta$ = $10^{-3}$
 only MMs with $m_M>10^{14}$ GeV can traverse the Earth.
Fig.~4 shows the accessible mass region for MMs of different velocities coming 
from above; for $g=g_D$ and $\beta=10^{-3}$ a MM must have 
$m_M >10^{10}$ GeV, $10^{6}$ 
GeV, $10^{5}$ GeV in order to reach the underground MACRO detector, the Earth 
surface, and a detector at 5230 m altitude, respectively [13].

\setcounter{figure}{3}

\begin{figure}[ht]
\vspace{-2cm}
\begin{center}
        \mbox{ \epsfysize=9.5cm
            \epsffile{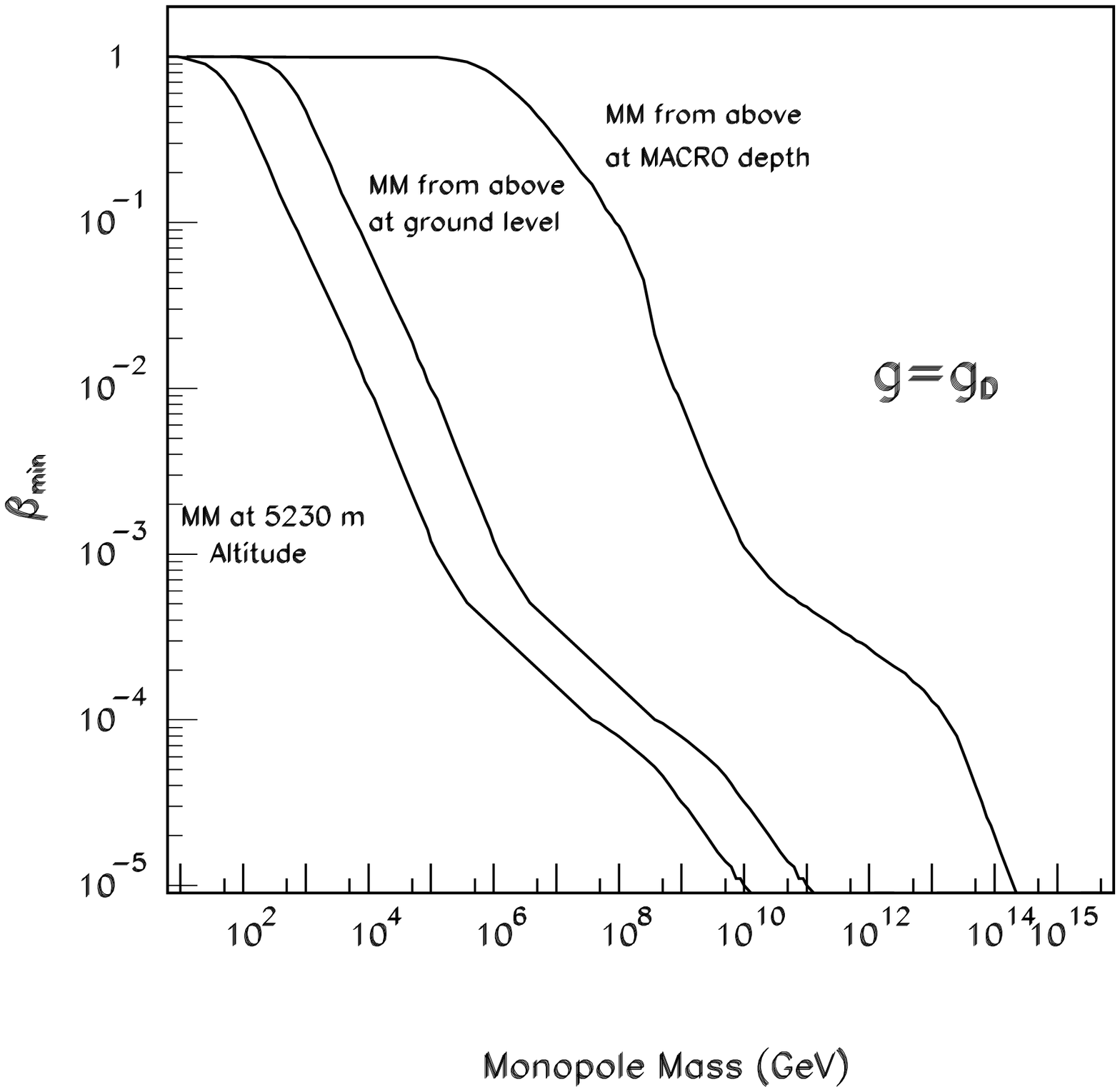}}
\vspace{-1cm}
\caption{\protect\small
Accessible region in the plane (mass, $\beta$) for monopoles with
magnetic charge $g= g_D$ from above for an experiment at an altitude of
5230 m, at sea level and
for an underground detector at the Gran Sasso Lab. (like the MACRO detector)
[13].}
\label{slim}
\end{center}
\end{figure}

\section{Experimental searches for supermassive GUT monopoles}

A flux of cosmic GUT supermassive magnetic monopoles may reach the Earth 
and may have done so for the whole life of the Earth. The velocity 
spectrum of these 
MMs  could be in the range $3\cdot 10^{-5} <\beta <0.1$.
Searches for such MMs have been performed with superconducting induction 
devices whose combined limit is at the level of  
$2\cdot$\units{10^{-14}~cm^{-2}~s^{-1}~sr^{-1}}, independent of $\beta$, see Fig.~5.
Direct searches were performed underground using 
scintillators, gaseous detectors and nuclear track detectors 
(mainly CR39) [24]. No 
monopoles have been detected, and the present 90\% C.L. flux 
limits are given in 
Fig.~5. Indirect searches were performed with old mica samples [25].

\begin{figure}[ht]
\vspace{-2cm}
\begin{center}
\mbox{ \epsfysize=10cm
            \epsffile{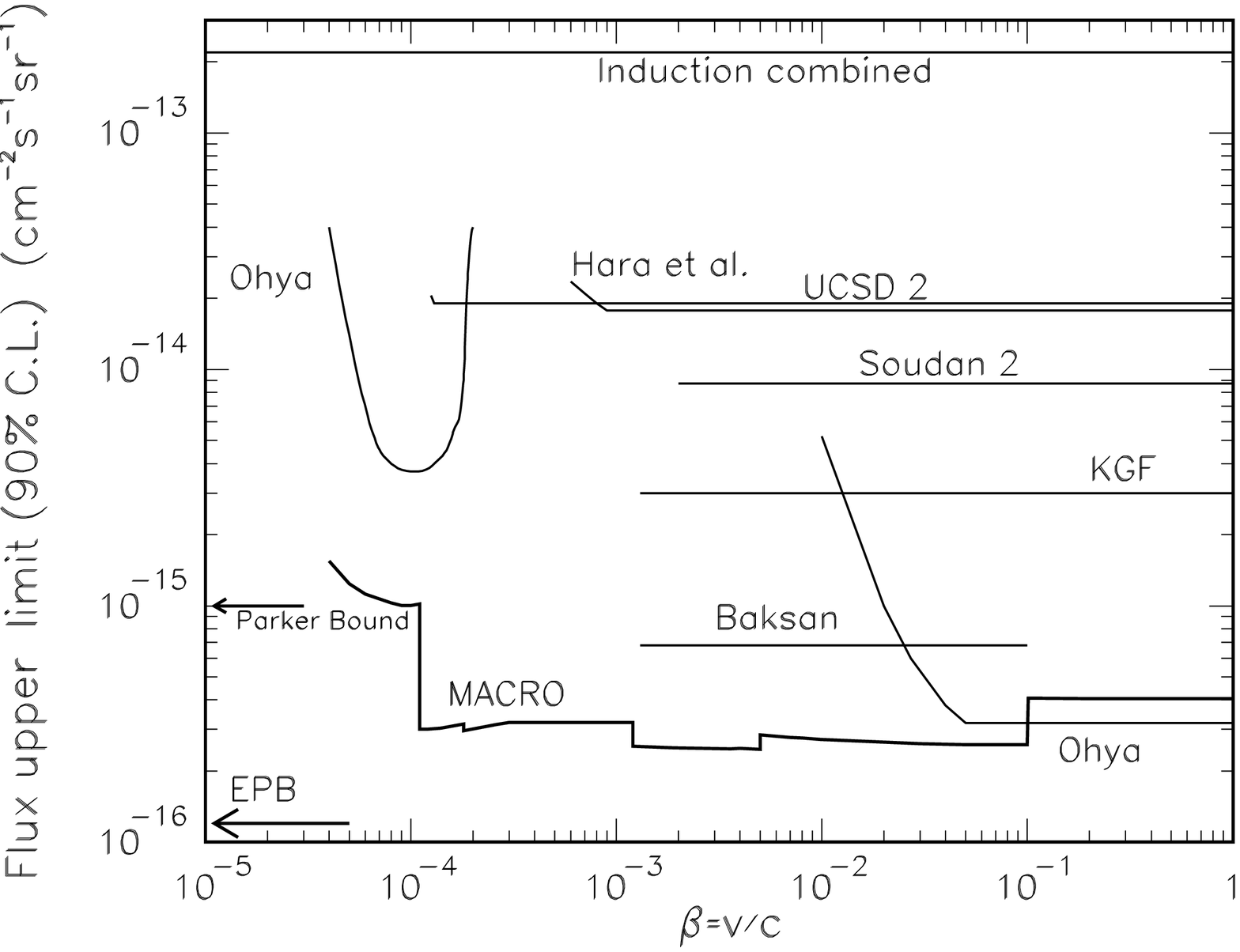}}
\caption{\protect\small
Compilation of 90\% C.L. of direct experimental upper limits on an isotropic
MM flux reaching detectors at the surface or
underground [24].}
\label{global}
\end{center}
\end{figure}

\section{Intermediate mass magnetic monopoles}
Magnetic monopoles with intermediate masses, $10^3~<m_M<~10^{12}$ GeV, have
been proposed by several authors, see f.e. Ref.~[5]. These monopoles could
be present in the cosmic radiation, but would not reach underground detectors,
see Fig.~4 [13]. Detectors at the Earth surface are capable to detect MMs 
coming from above if they have masses larger than $\sim 10^5-10^6$ GeV; lower
mass monopoles may be searched for with detectors located at high mountain
altitudes, or even higher, in balloons and in satellites. Few experimental
results are available [30]. An experiment has been proposed to search for
intermediate mass monopoles with nuclear track detectors at the Earth surface
and at high altitude [13]. 

\section{Magnetic monopole searches at accelerators}

Classical MMs have been searched for at every new accelerator up to masses of
about 1 TeV [15]. The main motivation is connected with the original proposal
 of Dirac, in 1931 [6]. With the advent of higher energy accelerators, in 
particular of the LHC, it should be possible to search for MMs with masses up
to about 7 TeV, a mass region which may explore the possibility of magnetic
monopoles connected with the electroweak unification scale [26]. A simple 
experiment using nuclear track detectors is proposed to be used with the LHC
 collider [14]. \par
Ginzburg and Schiller have proposed to analyze $\gamma-\gamma$ collisions
at high energy colliders, below monopole production threshold [27]. 
$\gamma-\gamma$ scattering should be enhanced due to the large coupling
of monopoles to photons. At the Tevatron the effect could be seen as pair 
production of photons  with energies of 200-400 GeV and roughly compensated
transverse momenta of 100-400 GeV/c. This search could explore monopole
masses of 1-2 TeV; at LHC the explored masses could be 7-19 TeV.

\section{Experimental searches for nuclearites }

Nuclearites (strange quark matter) should be aggregates of  $u$, $d$  and  $s$
  quarks in equal proportions and of electrons to ensure their
 electrical neutrality. They 
should be stable for all barion numbers in the range between ordinary 
nuclei and 
neutron stars ($A\sim 10^{57}$). They could be the ground state of QCD and 
could have 
been produced in the primordial Universe or in violent astrophysical processes.
 Nuclearites could contribute to the dark matter in the Universe. 
An upper limit 
on the nuclearite flux may be estimated assuming that  
$\Phi_{max.}~=~\rho_{ DM} v/(2 \pi M)$,
 where
 $ \rho _{ DM}
\simeq 10^{-24}$~\units{g cm^{-3}} is the local dark matter density;
 $M$ and $v$ are the mass and the velocity of
nuclearites, respectively [28,29].

The most relevant direct upper flux limits for nuclearites come from three
 large area experiments: the first two use the CR39 nuclear track detector; 
one experiment was 
performed at mountain altitude [30], the second at a depth of 
10$^4$~\units{g~cm^{-2}} in the Ohya 
mines [31]; the third experiment is the MACRO detector which uses also liquid 
scintillators besides nuclear track detectors [32]. Indirect experiments 
using old mica 
samples could yield  the lowest flux limits, but they are affected by inherent 
uncertainties [25]. 
Some exotic cosmic ray events were interpreted as due to incident nuclearites, 
for example the ``Centauro" events, the anomalous massive particles, etc. [33]. 
The 
interpretation of those possible signals are not unique and the used 
detectors are not redundant.

The main energy loss mechanism for nuclearites passing through matter is that 
of atomic collisions. While traversing a medium the nuclearites should 
displace the 
matter in their path by elastic or quasi-elastic collisions with the ambient 
atoms [24,28]. 
The energy loss rate is 
\begin{equation}
dE/dx = \sigma \rho v^2,
\end{equation}
where $\sigma$ is the nuclearite cross section,
$v$ its velocity and $\rho$ the mass density of the traversed medium.
For nuclearites with masses $M \geq
1.5$ \units{ng}
 the cross section may be approximated as:
\begin{equation}
\sigma \simeq \pi \cdot \left ( \frac{3M}{4 \pi \rho_N} \right ) ^{2/3}
\end{equation}

It is assumed that the density of strange quark matter 
is $\rho_N \simeq 3.5 \cdot 10^{14}$ \units{g~cm^{-3}}
somewhat larger than that of atomic nuclei. 
For typical galactic velocities, nuclearites 
with masses larger than 0.1 g  could traverse the Earth.
 For nuclearites of smaller masses the collisions are governed by their 
electronic 
clouds, yielding $\sigma \simeq \pi \cdot 10^{-16}$ \units
{cm^2}.
Most nuclearite searches were obtained as byproducts of superheavy magnetic 
monopole searches. The cosmic ray flux limits are therefore similar to 
those obtained for MMs.

In Fig.~6 is presented a compilation of limits for a flux of downgoing 
nuclearites compared with the dark matter limit, assuming a velocity at 
ground level  $\beta = v/c =2 \cdot 10^{-3}$. This speed corresponds to 
nuclearites of galactic or extragalactic 
origin. In the figure we extended the MACRO limit above the dark matter bound,
in order to show the transition to an isotropic flux for nuclearite masses 
larger than 0.1 g ($\sim 10^{23}$ GeV ).

\begin{center}
\vspace{0.5cm}
\begin{minipage}{.30\textwidth}
       \begin{center}
        \mbox{\hspace{-1cm}
            \epsfig{file=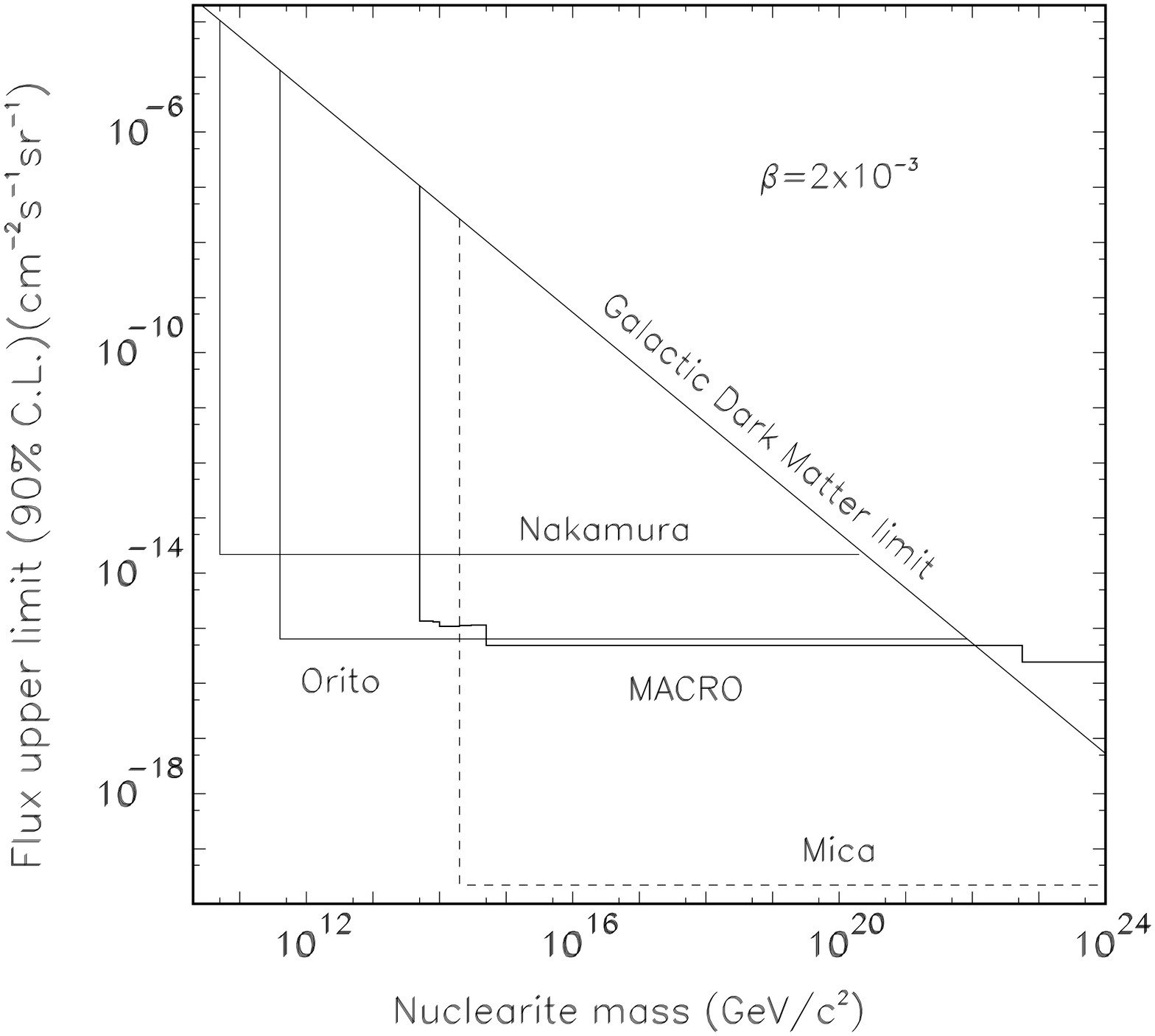,height=8.5truecm}
        }
\end{center}
\end{minipage}\hfill
\begin{minipage}{.40\textwidth}
{\small Figure 6: 90\% C.L. flux upper limits versus mass for nuclearites with
$\beta~=~2~\cdot~10^{-3}$ at ground level. These nuclearites
could have galactic or extragalactic origin. The limits are from 
  MACRO [24,~32],  from Refs.~[30] (``Nakamura''), [31] (``Orito'') and the indirect 
Mica limits of Ref.~[25].}
\end{minipage}
\end{center}

\section{Conclusions}
Comparing this lecture at the 5$^{th}$ School to that at the 
4$^{th}$ School [15] one 
notices the improvements made in this field during a 3-year period. The changes
concern some theoretical aspects [4,7,12], computation of the energy losses
in detectors [9] and in the Earth [10], new searches for MMs in the cosmic
radiation, see Fig.~5 [24], new methods to search for classical monopoles [29],
etc. [16]. In the fields related to that of MMs, because of similarity of 
detection techniques, one has theoretical extensions to Q-balls [17] and new
 limits on nuclearites, see Fig.~6 [32].\par

\vspace{1cm}

\noindent{\bf Acknowledgements.} We would like to acknowledge 
the cooperation of many colleagues of the MACRO
experiment, in particular M. Giorgini, T. Lari, G. Mandrioli, M. Ouchrif, 
V. Popa, P. Serra, M. Spurio, and others.


\end{document}